\documentclass[prb,aps,twocolumn,superscriptaddress]{revtex4-1}
\usepackage{amsmath}
\usepackage{latexsym}
\usepackage[english]{babel}
\usepackage{graphicx}
\DeclareGraphicsExtensions{.pdf,.png,.jpg, .jpeg, .eps}
\usepackage{bm}
\usepackage{array}
\usepackage{amssymb}
\usepackage{amsfonts}
\usepackage{multirow}
\usepackage{epstopdf}
\usepackage{color}
\usepackage{appendix}

\begin{document}

\title{Interfacial spin-orbit coupling driven enhancement of superconductivity by parallel magnetic field}

	\author{Zh. Devizorova}
	\affiliation{Moscow Institute of Physics and Technology, 141700 Dolgoprudny, Russia}
	\author{A.I. Buzdin}
	\affiliation{University Bordeaux, LOMA UMR-CNRS 5798, F-33405 Talence Cedex, France}
	\affiliation{World-Class Research Center “Digital biodesign and personalized healthcare”, Sechenov First Moscow State Medical University, Moscow 119991, Russia }
	
	\begin{abstract}
	It is commonly believed that time reversal symmetry breaking perturbations such as magnetic field or scattering on  magnetic impurities destruct superconductivity and suppress the critical temperature of the superconducting phase transition. However, in the recent experimental studies significant enhancement of superconducting critical temperature by parallel magnetic field was found for several thin superconducting systems [H.J. Gardner, et al., Nature Physics {\bf 7}, 895 (2011); T. Asaba, et al., Scientific Reports {\bf 8}, 1 (2018)]. Here we present a possible explanation of the observed phenomenon showing that combination of interfacial spin-orbit interaction and external magnetic field can cause the increase of the superconducting critical temperature of thin superconducting film in wide range of magnetic fields. We use the Ginzburg-Landau formalism taking into account Lifshitz invariant originated from Rashba spin-orbit interaction, which is also responsible for the superconducting diode effect. We also calculate the modification of the Little-Parks oscillations of the critical temperature of hollow superconducting cylinder at the presence of Rashba interfacial spin-orbit interaction.
	\end{abstract}
	
	\maketitle
Since Cooper pairs responsible for superconductivity are formed from time-reversal invariant electronic state, mechanisms that break this symmetry, including magnetic or exchange fields  and scattering on paramagnetic impurities, are expected to destroy superconducting state \cite{Maki_PR_1966,Ginzburg_1956_JETP,Abrikosov_1960}. There are two mechanism of this destruction: orbital \cite{Abrikosov_TheoryOfMetals} and paramagnetic \cite{Bulaevskii_AdvPhys_1985, Flouquet_PhysWorld_2002}. The orbital effect is related with a destruction of the Cooper pairs in magnetic field by Lorentz force, acting in opposite direction on the electrons of the pair, having the opposite momentum. On the other hand, the magnetic field tends to align spins of electron, which prevent superconducting singlet pairing with Cooper pairs formed from two electrons having opposite spin projections (paramagnetic mechanism). As a result, it is expected that the critical temperature of superconducting phase transition will be suppressed by applied magnetic field.

In the recent experimental studies the notable {\it enhancement} of critical temperature ($T_c$) by external magnetic field was found for several layered superconducting systems \cite{Gardner_NatPhys_2011, Asaba_SciRep_2018}, including ultrathin Pb films \cite{Gardner_NatPhys_2011}, two dimensional electron gas at LaAlO$_3$/SrTiO$_3$ interface \cite{Gardner_NatPhys_2011}, superconducting epitaxial thin film tungsten telluride \cite{Asaba_SciRep_2018}(WTe$_2$). For Pb films the enhancement of $T_c$ demonstrates systematic dependence on film thickness. Note that the reduction of pair-breaking effect due to polarization of paramagnetic impurities by applied magnetic field can causes similar enhancement of $T_c$ \cite{Kharitonov_JETPL_2005}. However, on experiment \cite{Gardner_NatPhys_2011} intentional addition of paramagnetic impurities (Cr) into Pb sample diminishes the effect. Thus, the effect is not due to polarization of paramagnetic impurities. The observed enhancement of critical temperature by the magnetic field can not be explained in the framework of standard Ginzburg-Landau (GL) theory. However, the analysis of this problem using the Eilenberger quasiclassical approach revealed the possibility of the field enhanced superconductivity \cite{Kogan_PRB_23, Kogan_PRB_86, Kogan_PRB_87}. The results obtained in Refs.[\onlinecite{Kogan_PRB_23, Kogan_PRB_86, Kogan_PRB_87}] show that this enhancement can be realised in thin superconducting film and disappears in the dirty limit. However, it is likely that the systems experimentally studied in Refs.[\onlinecite{Gardner_NatPhys_2011, Asaba_SciRep_2018}] are in dirty limit. Here we suggest an alternative explanation of the field enhanced superconductivity, based on the role of the spin-orbit coupling (SOC). Indeed, all systems studied in Refs.[\onlinecite{Gardner_NatPhys_2011, Asaba_SciRep_2018}] reveal the presence of SOC effects. Moreover, the enhancement of $T_c$ by applied magnetic field is more pronounced in Pb films, where SOC is larger, in comparison with LaAlO$_3$/SrTiO$_3$ sample \cite{Gardner_NatPhys_2011}. In addition, all samples are lack of inversion symmetry. Thus, one can expect that Rashba-type SOC \cite{Bychkov_1984, Rashba_1991}, which results from strong intrinsic spin-orbit interaction and lack of inversion symmetry in the sample, can play important role in the observed phenomenon.   
 
\begin{figure}[t!]
		\includegraphics[width=0.7\linewidth]{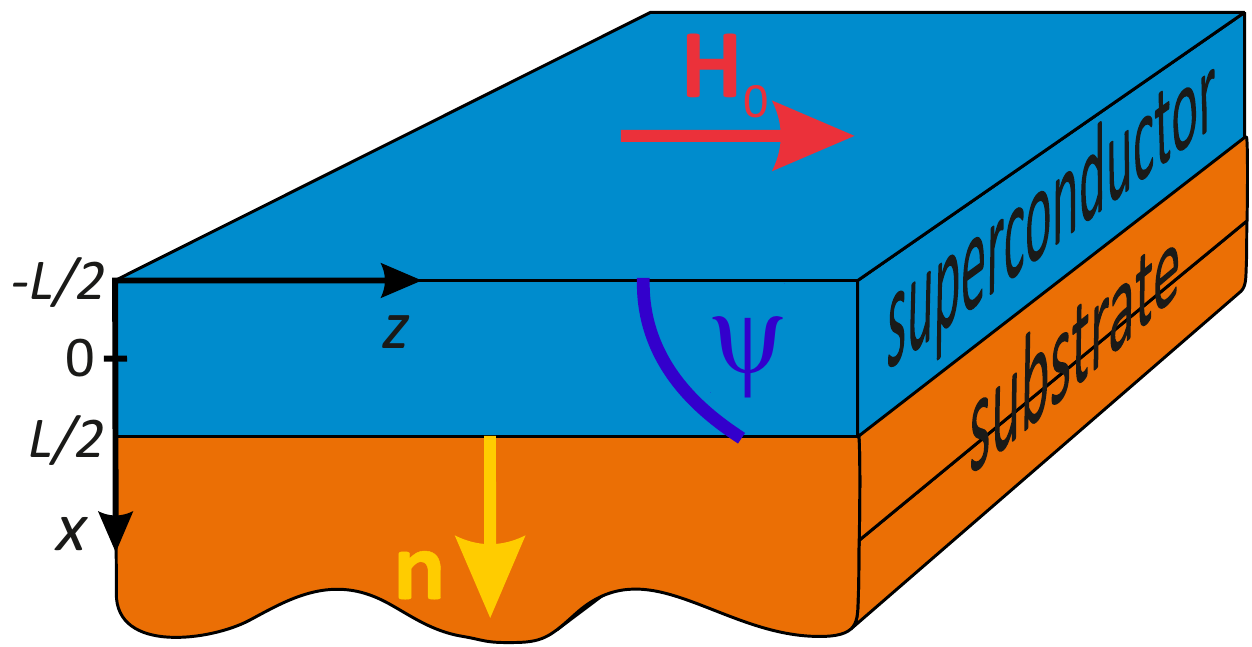}
		\caption{Sketch of a superconducting film lying on insulating substrate placed into the external magnetic field ${\bf H_0}=H_0{\bf e}_z$. The broken inversion symmetry results in Rashba spin-orbit coupling (SOC) on the bottom superconductor's interface.}\label{Fig:Film}
	\end{figure}

\begin{figure}[t!]
		\includegraphics[width=0.8\linewidth]{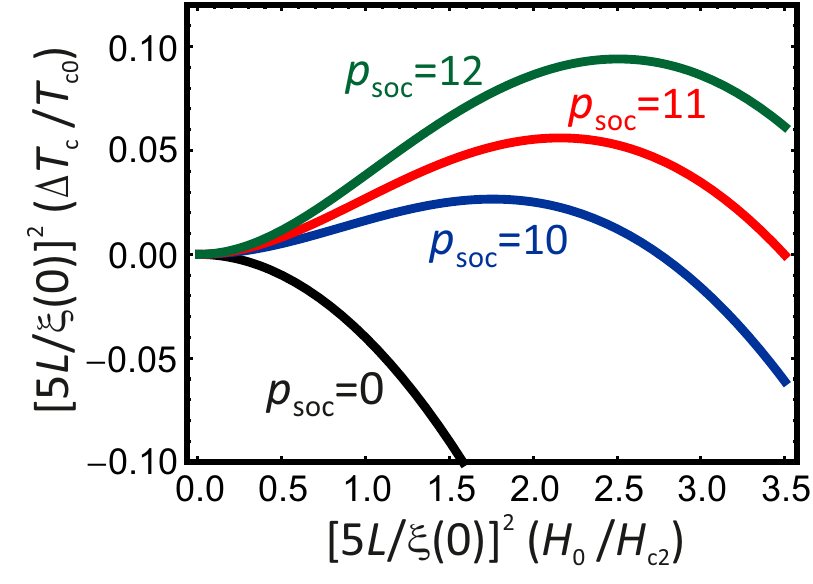}
		\caption{The superconducting critical temperature $T_c$ of a thin ($L=0.2\xi \sqrt{b}$) superconducting film with interfacial Rashba SOC vs. the external magnetic field $H_0$. The parameter $p_{soc}=s_{soc}\left[2\xi(0)/L\right]^2$ is responsible for the enhancement of $T_c$ due to SOC, since $s_{soc}\sim k_{soc}/k_F$ is material dependent SOC constant. Here $\eta=8$ (this parameter describes the suppression of $T_c$ caused by orbital and paramagnetic effects).} \label{Fig:ThinFilm_Tc_H}
	\end{figure}

\begin{figure}[t!]
		\includegraphics[width=0.4\linewidth]{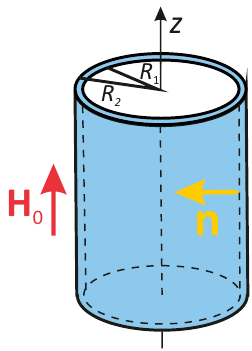}
		\caption{Sketch of a hollow superconducting cylinder  placed into magnetic field. The broken inversion symmetry at the internal surface results in the emergence of Rashba SOC.}\label{Fig:Cylinder}
	\end{figure}

\begin{figure}[t!]
		\includegraphics[width=0.8\linewidth]{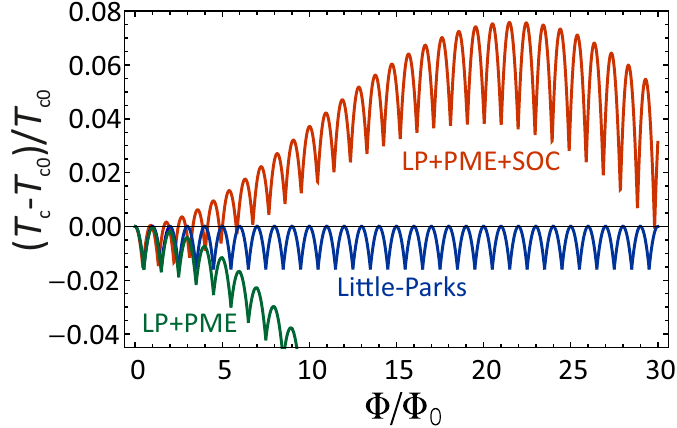}
		\caption{The superconducting critical temperature $T_c$ of a thin-walled cylinder as a function of the external magnetic flux $\Phi$. Here Rashba SOC parameter $s_{soc}=0.1$ for the red curve, while SOC is absent for the blue and green ones. The  paramagnetic parameter is $\alpha_p=0.06$ for the red and green curves. The blue curve demonstrates Little-Parks (LP) oscillations without both paramagnetic effect (PME) and SOC. For all curves $R_1/\xi(0)=4$, $\Delta R/\xi(0)=0.2$ and $b\sim 1$.}\label{Fig:Tc_Phi}
	\end{figure}

In the present paper we show that external magnetic field can cause the {\it enhancement} of $T_c$ (see Fig.\ref{Fig:ThinFilm_Tc_H}) of thin superconducting film with interfacial Rashba SOC, schematically shown in Fig.\ref{Fig:Film}. We use phenomenological GL formalism with additional term named Lifshitz invariant, which is proportional to SOC parameter and magnetic or exchange field. We show that this extra term leads to the nonzero derivative of the order parameter on the interface with broken inversion symmetry. In semi-infinite superconductor (S) placed in contact with thin ferromagnetic insulator (F) such non-zero derivative can cause the local enhancement of the order parameter near the S/F interface \cite{Mironov_PRL_2017}. Moreover, if the thickness of S layer in such S/F hybrid structure is of the order of superconducting coherence length, one can expect the enhancement of its critical temperature \cite{Devizorova_PRB_21}. Here we show, that additional F layer is not required to have Lifshitz invariant induced enhancement of $T_c$ and it can be realized in thin S film with Rashba SOC by applying external magnetic field (see Fig.\ref{Fig:ThinFilm_Tc_H}). To describe the bend in $T_c(H)$ dependence in higher magnetic field, we take into account the third over gradients term in Lifshitz invariant, which is also responsible for the superconducting diode effect widely discussed in literature \cite{Yuan_PNAS_2022, Ando_Nature_2020, Daido_PRL_2022, He_NJP_2022, Baumgartner_NatNano_2022, Nadeem_NatRevPhys_23}. We also calculate $T_c$ for a thin-walled superconducting cylinder with interfacial Rashba SOC, schematically shown in Fig.\ref{Fig:Cylinder}. It is well-known, that without SOC $T_c$ of such a system is suppressed by external magnetic field and exhibits Little-Parks oscillations as a function of external magnetic flux \cite{Parks_PR_1964}. Here we show that Rashba SOC at the surface of the cylinder can give rise to the total enhancement, but not suppression of its $T_c$, superimposed by Little-Parks like oscillations (see Fig.\ref{Fig:Tc_Phi}).

The GL functional of a superconductor with interfacial Rashba-type SOC placed into external magnetic field reads as \cite{Mironov_PRL_2017,Samokhin_PRB_2004, Kaur_PRL_2005,Yuan_PNAS_2022}:

\begin{multline}
		\label{Eq:G}
		G = \int dV \Biggl(\alpha |\psi|^2 + \frac{\beta}{2}|\psi|^4 + \frac{\hbar^2}{4m}|\hat{\boldsymbol{D}}\psi|^2 +  \frac{(\bm{B} - \bm{H_0})^2}{8\pi} + \\+\epsilon_{so}({\bf r})\boldsymbol{n}\times \boldsymbol{B}\biggl[\psi^*\hat{\boldsymbol{D}}(1-b\xi ^2\hat{\boldsymbol{D}}^2)\psi + \\+\psi\left(\hat{\boldsymbol{D}}(1-b\xi^2\hat{\boldsymbol{D}}^2)\psi\right)^*\biggr] +\frac{(\chi_n-\chi_s)H_0^2}{2}\Biggr).
\end{multline}
 Here $\psi$ is the superconducting order parameter, $\alpha = a(T-T_{c0})$ and $\beta$ are the standard GL coefficients, $T_{c0}$ is the critical temperature of a bulk superconductor, $\hat{\boldsymbol{D}} = -i\boldsymbol{\nabla} + \frac{2e}{c\hbar}\boldsymbol{A}$, ${\bf A}$ is the vector-potential, $\bf{B}=\boldsymbol{\nabla} \times \bf{A}$ is the magnetic field induction, $\bf{H}_0$ is the external magnetic field, $\boldsymbol{n}$ is the unit vector directed as the external normal to the superconductor's interface with Rashba SOC, $\epsilon_{so}({\bf r})$ is the spin-orbit constant, which is nonzero only in the atomically thin area of the width $l_{so}$ near the interface with Rashba SOC, $\xi$ is superconducting coherence length [$\xi=\xi_0=0.18 \hbar v_F/T_{c0}$ in pure limit and $\xi=\sqrt{D_s/(2\pi T_{co})}$ in dirty limit, where $D_s$ is diffusion coefficient], $b \approx 1$ is numerical constant ($b=1.13$ in pure limit \cite{Yuan_PNAS_2022}). The last term in Eq.\ref{Eq:G} describes paramagnetic effect since $\chi_n$ and $\chi_s$ are the paramagnetic susceptibilities of normal and superconducting states, respectively. 

We calculate the superconducting critical temperature $T_c$ of the superconducting film with the thickness $L$ lying on insulating substrate, placed into the external magnetic field ${\bf H}_0$ (see Fig.\ref{Fig:Film}). Due to lack of inversion symmetry in the structure, interfacial Rashba SOC appears at the superconductor/substrate interface. Thus, the structure is described by GL functional Eq.(\ref{Eq:G}). Without loss of generality, we assume that the $x$-axis coincides with ${\bf n}$, the $z$-axis is directed along the external magnetic field $\boldsymbol{H}_0=H_0 \boldsymbol{e}_z$ and the vector potential is along the $y$-axis $\boldsymbol{A}=A_y \boldsymbol{e}_y$. Superconducting film occupies the region $-L/2<x<L/2$. 

Since near superconducting phase transition the order parameter is small, we write down the GL equations in the lowest order over $\psi$ and neglect the contribution of screening currents to the magnetic field. Thus, we take $\boldsymbol{B} \approx \boldsymbol{H}_0$ and $A_y=H_0 x$. Since the vector-potential is directed along the $y$ axis, we can choose the ansatz for the order parameter in the form: $\psi =e^{ik_y y}\psi(x)$. 

The GL equation for the order parameter and the boundary condition (BC) on the top surface have the standard form and read as: $ \partial_x\psi(x)|_{x=-L/2}=0$,

\begin{equation}
    \alpha \psi(x)-\frac{\hbar^2}{4m}\partial_{xx}\psi(x)+\frac{\left(\hbar k_y+2eH_0x/c\right)^2}{4m}\psi(x)=0.
\end{equation}
Note that the above GL equation is applicable only for thick films with $L>\xi$. In such the case near $T_c$ we can neglect the paramagnetic contribution since it is small compared to the orbital one. 

The BC at the bottom interface is modified by interfacial Rashba SOC. We derive it minimizing the GL functional Eq.(\ref{Eq:G}) and show, that the derivative of the order parameter on the bottom surface is nonzero due to the presence of both magnetic field and Rashba SOC:

\begin{multline}
   \partial_x \psi(x)|_{x=L/2}=s_{soc}\frac{H_0}{H_{c2}}\left(k_y+\frac{eH_0L}{c\hbar}\right)\times \\ \times \biggl[1-b \xi^2 \left(k_y+\frac{eH_0L}{c\hbar }\right)^2\biggr]\psi(L/2).
\end{multline}
Here the dimensionless parameter $s_{soc}=8k_{soc}l_{so}$ is material dependent and describes the strength of SOC, $k_{soc}= m \epsilon_{so}H_{c2}/\hbar^2$ is the momentum determined by SOC constant $\epsilon_{so}$ and the upper critical field $H_{c2}=\Phi_0/[2\pi \xi^2(0)]$, where $\Phi_0=\pi \hbar c/e$ is the superconducting flux quantum and $\xi(T)=\sqrt{\hbar^2/[4m a(T_{c0}-T)]}$ is the temperature dependent superconducting coherence length. Note that $s_{soc}\sim k_{soc}/k_F$, where $k_F$ is Fermi momentum, is hardly large than unity. 

 Following the procedure described in Ref.~[\onlinecite{Saint_James}] we introduce the dimensionless coordinate $X=2x/L$, the modulation vector $K_y=k_yL/2$, the parameters $\tilde{H}_0=eH_0L^2/(2\hbar c)=[L/2\xi(0)]^2(H_0/H_{c2})$, $\epsilon=-m\alpha L^2/\hbar^2$  and rewrite the GL equation in the following form:
 
\begin{equation}
    \partial_{XX}\psi(X)+(K_y+\tilde{H}_0X)^2\psi(X)=\epsilon \psi(X).
\end{equation}

At the same time, the BSs read: $\partial_{X} \psi(-1)=0$, 

\begin{multline}
    \left. \left(\frac{\partial_X \psi}{\psi}\right) \right |_{X=1}=p_{soc}\tilde{H}_0(K_y+\tilde{H}_0)\times \\ \times \biggl[1-b\left(\frac{2\xi}{L}\right)^2(K_y+\tilde{H}_0)^2\biggr].
\end{multline}
Here the dimensionless parameter $p_{soc}=s_{soc}\left[2\xi(0)/L\right]^2$ is determined by the material dependent SOC constant $s_{soc}$ and the film thickness $L$. 

Next it is useful to introduce a new variable $t=\sqrt{2|\tilde{H}_0|}(X+K_y/\tilde{H}_0)$. The resulting GL equation and the BCs are the following:

\begin{equation}
\label{Weber}
     -\partial_{tt}\psi(t)+\frac{1}{4}t^2\psi(t)=\frac{\epsilon}{2|\tilde{H}_0|}\psi(t),
\end{equation}

\begin{equation}
        \partial_t \psi|_{\sqrt{2|\tilde{H}_0|}(-1+K_y/\tilde{H}_0)}=0,
\end{equation}

\begin{multline}
 \left.\left(\frac{\partial_t \psi}{\psi}\right)\right|_{\sqrt{2|\tilde{H}_0|}(1+K_y/\tilde{H}_0)}=\frac{p_{soc}\tilde{H}_0(K_y+\tilde{H}_0)}{\sqrt{2|\tilde{H}_0|}}\times \\ \times \biggl[1-b\left(\frac{2\xi}{L}\right)^2(K_y+\tilde{H}_0)^2\biggr].
\end{multline}
   
The solution of Eq.(\ref{Weber}) has the form of the linear combination of the Weber functions $\psi(t)=A_{\nu}D_{\nu}(t)+B_{\nu}D_{\nu}(-t)$, where $2\nu+1=\epsilon/|\tilde{H}_0|$. Substituting it into the BCs we obtain the following equation, which implicitly defines the function $T_c(H_0)$:

\begin{multline}
\label{nu}
    D'_{\nu}(t_+) D'_{\nu}(-t_-)-D'_{\nu}(-t_+) D'_{\nu}(t_-)=\\=\frac{p_{soc}\tilde{H}_0(K_y+\tilde{H}_0)}{\sqrt{2|\tilde{H}_0|}}  \biggl[1-b\left(\frac{2\xi}{L}\right)^2(K_y+\tilde{H}_0)^2\biggr] \times\\ \times\biggl[D_{\nu}(t_+) D'_{\nu}(-t_-)+D_{\nu}(-t_+) D'_{\nu}(t_-)\biggr],
\end{multline}
where $t_{\pm}=\sqrt{2|\tilde{H}_0|}(\pm 1 +K_y/\tilde{H}_0)$.

\begin{figure}[t!]
		\includegraphics[width=0.8\linewidth]{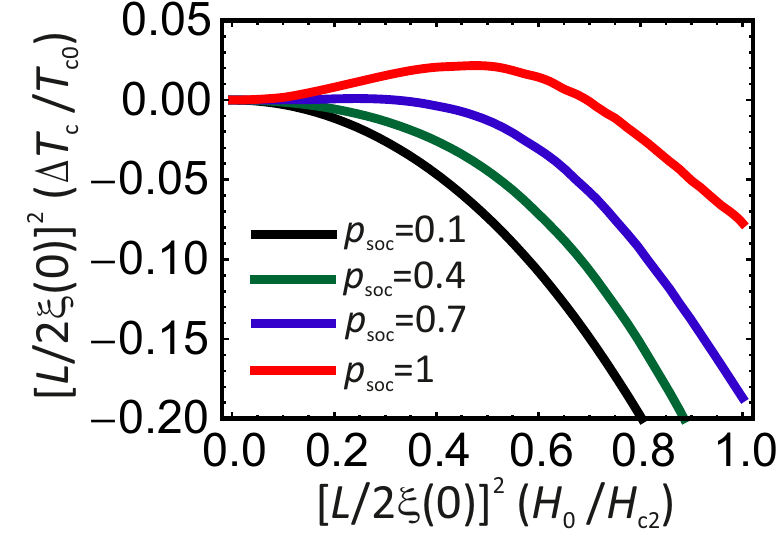}
		\caption{The superconducting critical temperature $T_c$ of a thick ($L=2\sqrt{b}\xi$) superconducting film with interfacial Rashba SOC vs. the external magnetic field $H_0$.}\label{Fig:Tc_H}
	\end{figure}

The maximal temperature, at which the film is in superconducting state at fixed magnetic field, corresponds to the minimal value of $\nu$, satisfying Eq.(\ref{nu}). At fixed $K_y$ and $\tilde{H}_0$ this equation has infinite but discrete number of solutions for $\nu$, and we find the minimal one. Then we minimize it with respect to $K_y$ and find the minimal value $\nu_0$ for fixed $\tilde{H}_0$. As a result, we obtain the dependence $\epsilon(\tilde{H}_0)$ in the form $\epsilon=(2\nu_0+1)|\tilde{H}_0|$. Since $\epsilon=-[L/2\xi(0)]^2 (T_c-T_{c0})/T_{c0}$ and $\tilde{H}_0=[L/2\xi(0)]^2 (H_0/H_{c2})$, we plot the maximal temperature when the film is in superconducting state ($T_c$) at fixed external magnetic field vs. this field $H_0$ (see. Fig.\ref{Fig:Tc_H}).

Our calculations shows, that interfacial Rashba SOC increases the superconducting critical temperature $T_c$ of a superconducting film in comparison with a film without SOC (see. Fig.\ref{Fig:Tc_H}). However, for reasonable choise of SOC parameter, in thick superconducting film with $L \gtrsim \xi$, for which three-dimensional GL equation is applicable, total enhancement of $T_c$ is hardly realised. Indeed, at small magnetic fields the increase in $T_c$ due to Lifshitz invariant is of the order of $(\Delta T_c/T_{c0}) \approx (s_{soc}/2)^2[\xi(0)/L]^2[H_0/H_{c2}]^2$. At the same time, the decrease of $T_c$ caused by orbital effect is proportional to $(\Delta T_c/T_{c0}) \approx -(1/6)[L/\xi(0)]^2[H_0/H_{c2}]^2$. As a result, to have total enhancement of $T_c$ at small magnetic field we need $p_{soc} \gtrsim 1$. Thus, for $L=2\sqrt{b}\xi$ the threshold value of SOC parameter to have total enhancement of $T_c$ is of the order of $s_{soc} \sim 1$. This estimates is confirmed by more careful analysis, since for the film with the thickness $L=2\sqrt{b}\xi$ the $T_c$ is enhanced at $p_{soc}=1$, see. Fig.\ref{Fig:Tc_H}. For thicker films we need $s_{soc} \gg 1$, which is hardly realised in real materials. Thus, the enhancement of $T_c$ can be possible only in thin superconducting films with $L \ll \xi$. Below we analyse such the case.

In thin superconducting films with the thickness $L \ll \xi$ the superconducting order parameter does not depend on $x$-coordinate and can be represented as $\psi=\psi_0 e^{ik_yy}$. In such the case the effective two-dimensional GL functional near the phase transition has the following form:

\begin{multline}
\label{G_thin}
        \frac{G}{S}=\frac{\hbar^2 \psi_0^2}{mL}\biggl[-\epsilon+K_y^2+\frac{\eta}{2}\tilde{H}_0^2-\frac{1}{2}p_{soc}\tilde{H}_0 K_y \times \\ \times \left(1-(2\sqrt{b}\xi/L)^2K_y^2 \right)\biggr].
\end{multline}
Here $\eta=\{2/3+[2\xi(0)/L]^2\alpha_p\}$, where parameter $\alpha_p \propto [2\mu_B H_{c2}/(\pi T_{c0})]^2$ is responsible for the paramagnetic effect. Note that strong spin-orbit scattering can suppress paramagnetic mechanism of superconductivity destruction \cite{Werthamer_PR_1966, Bulaevskii_PRB_86}. In dirty samples with normal scattering is much more frequent, than spin-orbit one, the paramagnetic parameter can be estimated as $\alpha_p \approx [2\mu_B H_{c2}/(\pi T_{c0})]^2(1+\lambda_{scat})^{-1}$, where $\lambda_{scat}=2/(3\pi T_{c0}\tau_{so})$ and $\tau_{so}$ is spin-orbit scattering time \cite{Werthamer_PR_1966}. If spin-orbit scattering is strong, i.e. $\pi T_{c0}\tau_{so} \lesssim 1$, the paramagnetic parameter $\alpha_p$ can be significantly reduced.

The optimal modulation vector minimizing the GL functional Eq.(\ref{G_thin}) has the form:

\begin{equation}
\label{Ky}
    K_y=\frac{2}{3 p_{soc} \tilde{H}_0 A^2}\left(\sqrt{1+\frac{3}{4}p_{soc}^2 \tilde{H}_0^2 A^2}-1\right),
\end{equation}
where $A=2\sqrt{b}\xi/L$. Since at superconducting phase transition $G=0$, the change in the critical temperature $\Delta T_c=(T_c-T_{c0})$ caused by magnetic field and spin-orbit coupling is the following:

\begin{equation}
    \frac{\Delta T_c}{T_{c0}}=\frac{4\xi(0)^2}{L^2}\biggl[-K_y^2-\frac{\eta}{2}\tilde{H}_0^2+\frac{1}{2}p_{soc}\tilde{H}_0K_y(1-A^2K_y^2)\biggr],
\end{equation}
where $K_y$ is determined by Eq.(\ref{Ky}).

The typical dependencies $\Delta T_c(H_0)$ for different SOC parameters are shown in Fig.\ref{Fig:ThinFilm_Tc_H}. Note that for $L\sim 0.2\xi$ the parameter $p_{soc}=12$ correspond to SOC constant $s_{soc} \sim 0.12$, which is reasonable value. We also assume strong spin-orbit scattering in the system, which decrease the paramagnetic contribution to $\Delta T_c$.

Our calculations show, that for reasonable choice of parameters the superconducting critical temperature $T_c$ of a thin ($L \ll \xi$) superconducting film with interfacial Rashba SOC can be enhanced by external magnetic field in wide range of the fields (see. Fig.\ref{Fig:ThinFilm_Tc_H}). This enhancement is caused by Lifshitz invariant in the GL functional Eq.(\ref{Eq:G}). The results (see. Fig.\ref{Fig:ThinFilm_Tc_H}) are in qualitative agreement with the experimental data obtained in Refs.[\onlinecite{Gardner_NatPhys_2011, Asaba_SciRep_2018}]. Note that we expect spin-orbit scattering to be strong in all systems experimentally studied in Refs.[\onlinecite{{Gardner_NatPhys_2011, Asaba_SciRep_2018}}].

Interfacial Rashba SOC can also modify Little-Parks oscillations of $T_c$ (see Fig.\ref{Fig:Tc_Phi}). To show this, let us consider a hollow superconducting cylinder having internal (external) radius $R_1$ ($R_2$) with broken inversion symmetry at the internal surface, resulting in Rashba SOC, placed into the external magnetic field ${\bf H}_0=H_0 {\bf e}_z$ (see Fig.\ref{Fig:Cylinder}). We assume the cylinder to be thin-walled, i.e. $\Delta R=(R_2-R_1)\ll \xi(0)$. In such the case the order parameter in cylindrical coordinate system $(r,\varphi,z)$ reads as $\psi({\bf r})=\psi_0 \exp{(il\varphi)}$, where $l$ is integer angular momentum. As a result, the relative change of the critical temperature $\tau=(T_c-T_{c0})/T_{c0}$ is the following:

\begin{multline}
\label{Eq:Tc_Phi}
    \tau=-\frac{(l+\mu)^2}{\rho^2}-\frac{2\alpha_p}{\rho^4}\mu^2 - \frac{2s_{soc}}{\rho^4 \delta}\mu(l+\mu) \times \\ \times \left[1-\tilde{A}^2(l+\mu)^2\right],
\end{multline}
Here $\mu=\Phi/\Phi_0$, where $\Phi=\pi H_0R_1^2$ is the external magnetic flux through the cylinder, $\rho=R_1/\xi(0)$, $\delta=\Delta R/R_1$ and $\tilde{A}=\sqrt{b}\xi/R_1$.

The first term in Eq.(\ref{Eq:Tc_Phi}) describes usual Little-Parks oscillations \cite{Parks_PR_1964}. The second one is responsible for the suppression of $T_c$ due to the paramagnetic effect. In contrary, the last term in Eq.(\ref{Eq:Tc_Phi}), caused by the combination of interfacial Rashba SOC and external magnetic field, leads to the enhancement of $T_c$. As a result, the envelope function of the $T_c(\Phi)$ dependence demonstrates nonmonotonic behaviour: it grows initially, reaches the maximum and then declines. This nonmonotonic envelope is superimposed by Little-Parks like oscillations (see Fig.\ref{Fig:Tc_Phi}). As a result, for reasonable choice of parameters the $T_c$ of a hollow thin-walled cylinder could be enhanced, but not suppressed, by external magnetic flux. 

In conclusion, we show that the critical temperature of a thin superconducting film with broken inversion symmetry and strong SOC in the bulk can be enhanced by external magnetic field (see Fig.\ref{Fig:ThinFilm_Tc_H}). This effect is caused by Lifshitz invariant in GL functional, which results from combination of external magnetic field and interfacial Rashba SOC, appearing, on its tern, due to the lack of structural inversion symmetry combined with bulk SOC. The results are in qualitative agreement with the experimental data \cite{Gardner_NatPhys_2011, Asaba_SciRep_2018}. The same mechanism can cause the modification of Little-Parks oscillations in cylindrical geometry (see Fig.\ref{Fig:Tc_Phi}).

The authors express their gratitude to V.G. Kogan for his valuable correspondence. This work is supported by Russian Science Foundation (grant \#22-72-00083) in part related to the calculation of the critical field of thin film and Ministry of Science and Higher Education of the Russian Federation within the framework of state funding for the creation and development of World-Class Research Center (WCRC) “Digital biodesign and personalized healthcare” (grant \#075-15-2022-304) in part related to the calculation of the influence of the spin-orbit interaction on the Little-Parks oscillations. Zh.D. acknowledges the support from ”Basis” foundation.

\bibliography{bib}

\end{document}